\documentclass[12pt]{article}
\usepackage[dvips]{graphicx}
\usepackage{amssymb}
\usepackage{amsmath,amsthm}

\usepackage{bm}
\bmdefine{\Bzero}{0}
\bmdefine{\Bone}{1}

\def\Bx{{\bf x}}

\def\Bd{{\bf d}}
\def\Bn{{\bf n}}

\def\BA{{\rm A}}
\def\BB{{\rm B}}
\def\BC{{\rm C}}
\def\BD{{\rm D}}
\def\BE{{\rm E}}
\def\BF{{\rm F}}

\title{Statistical testing procedure for the interaction
effects of several controllable factors in two-valued input-output
systems}
\author{
Satoshi Aoki\\
Department of Mathematics and Computer Science\\
Kagoshima University\\
and\\
Masami Miyakawa\\
Department of Industrial Engineering and Management\\
Tokyo Institute of Technology
}
\date{June, 2007}
\begin{document}
\maketitle
\begin{abstract}
Suppose several two-valued input-output systems are designed by setting
the levels of several controllable factors. For this situation, Taguchi method
has proposed to assign the controllable factors to the orthogonal
array 
and use ANOVA model for the standardized SN ratio, which is a natural 
measure for evaluating the performance of each input-output system. 
Though this procedure is simple and useful in application indeed, 
the result can be unreliable when the estimated standard errors of the
 standardized
SN ratios are unbalanced. In this paper, we treat the data arising 
from the full factorial
or fractional factorial designs of several controllable factors as 
the frequencies of high-dimensional contingency tables, and propose a general
testing procedure for the main effects or the interaction effects of the 
controllable factors.

\ \\
{\it Keywords}: 
Confoundings, 
Contingency tables,
Controllable factors,
Covariate matrix,
Generalized linear models, 
Hierarchical models,
Fractional factorial designs, 
Full factorial designs, 
Standardized SN ratio, 
Sufficient statistics,
Two-valued input-output systems.

\end{abstract}

\section{Introduction}
In this paper, we consider evaluating performance of
several two-valued input-output systems. 
Before introducing our motivated problem, first we give a typical
example of a {\it single} two-valued input-output system and review
the measure for evaluating its performance.
Suppose a vending machine judges an inserted coin as fair coin
or false coin. 
In this system, the input is the {\it true state} of the
inserted coin, \{fair, false\}, whereas the output is the {\it judged
state} of the coin, \{fair, false\}. For evaluating performance of
this machine, prepare $n_{1}$ fair coins and $n_{2}$ false coins and
insert them to the machine. The result of the judgment by this machine
is summarized as a $2\times 2$ contingency table as follows.
\begin{center}
\begin{tabular}{c|cc|c}
 & $y = 1$ & $y = 2$ & total\\ \hline
$M = 1$ & $n_{11}$ & $n_{12}$ & $n_{1}$\\ 
$M = 2$ & $n_{21}$ & $n_{22}$ & $n_{2}$\\ \hline
\end{tabular}
\end{center}
In the above table, two-valued signal $M$ is the true
state of the coin, and $y$ is the judged state by 
this machine (1: fair, 2: false). A natural statistical model for this 
experiment is two independent
binomial distributions. We introduce random variables, $N_{11}, N_{21}$, 
and parameters $p_{ij},\ i,j = 1,2$, 
and write $N_{i1} \sim Bin(n_i, p_{i1}),\ i = 1,2$.
$p_{ij}$ is the probability that a coin of true state $i$ is judged as a
state $j$. Here $p_{i1} + p_{i2} = 1, i = 1,2$ holds.
For this type of the systems, 
the error of judging a fair coin as false
(type I error) and the error of judging a false coin as fair (type II
error) are in the trade-off relation. Therefore we have to take into
account 
the two types of errors for evaluating the performance of the system.
In Taguchi method, a standardized Signal-Noise (SN) ratio, 
\begin{equation}
\hat{\eta} = -\log\left[
\frac{1}{(1 - 2\hat{p}_0)^2} - 1
\right],
\label{eqn:SNratio}
\end{equation}
is proposed as a measure to quantitatively evaluate performance of 
this type of system, where
\[
\hat{p}_0 = \frac{1}{1 + \sqrt{\hat{\theta}}}
\]
is the estimated common error rate, and
\begin{equation}
\hat{\theta} = \frac{n_{11}n_{22}}{n_{12}n_{21}}
\label{eqn:sample-odds-ratio}
\end{equation}
is the sample odds ratio. See Taguchi (1987) and Taguchi (1991).
Note that the standardized SN 
ratio (\ref{eqn:SNratio}) is a function of the sample odds ratio.

Now we suppose several two-valued input-output systems are designed
by setting the levels of several controllable factors, which is
a situation that we mainly consider in this paper.
Table \ref{tbl:example-miyakawa-7.3} is an example of such
situations shown in Section 7.3.2 of Miyakawa (2006).
In this experiment, 40 normal products and 20 failure products
are judges whether normal or failure, by a testing
inspection machine, which makes use of the leaking helium gas, 
in the casing process of a compressor. 
There are three controllable factors which are considered to have
influence on performance of the inspection, each of which is settled
on one of the two levels.
\begin{table*}[htbp]
\begin{center}
\caption{The result of an experiment judging 40 good products 
and 20 bad products as good or bad by a testing inspection machine
shown in Miyakawa (2006)}
\begin{tabular}{|c|ccc|rrrr|c|r|}\hline
No. & A & B & C & \multicolumn{1}{|c}{$n_{11}$} &
 \multicolumn{1}{c}{$n_{12}$} & \multicolumn{1}{c}{$n_{21}$} &
 \multicolumn{1}{c|}{$n_{22}$} & $\hat{p_0}$ &
 \multicolumn{1}{|c|}{$\hat{\eta}$}\\ \hline
1 & 1 & 1 & 1 & 27 & 13 & 11 & 9 & 0.435 & -17.689\\
2 & 1 & 1 & 2 & 25 & 15 & 3 & 17 & 0.259 & -5.169\\
3 & 1 & 2 & 1 & 17 & 23 & 8 & 12 & 0.489 & -32.873\\
4 & 1 & 2 & 2 & 15 & 25 & 2 & 18 & 0.320 & -8.295\\
5 & 2 & 1 & 1 & 40 & 0 & 12 & 8 & 0.119 & 1.427\\
6 & 2 & 1 & 2 & 38 & 2 & 10 & 10 & 0.203 & -2.638\\
7 & 2 & 2 & 1 & 40 & 0 & 11 & 9 & 0.109 & 1.974\\
8 & 2 & 2 & 2 & 37 & 3 & 10 & 10 & 0.234 & -4.037\\ \hline
\end{tabular}
\label{tbl:example-miyakawa-7.3}
\end{center}
\end{table*}
In Table \ref{tbl:example-miyakawa-7.3}, the meaning of $n_{ij}$ is
the same to the previous example. Note that the sample odds ratio 
(\ref{eqn:sample-odds-ratio}) is
modified as
\begin{equation}
\hat{\theta} = \frac{(n_{11}+0.5)(n_{22}+0.5)}{(n_{12}+0.5)(n_{21}+0.5)},
\label{eqn:modification}
\end{equation}
since some zeros are included in the table. For this type of the data,
Taguchi method has proposed to use ANOVA model for the standardized SN
ratio. For the data of Table \ref{tbl:example-miyakawa-7.3}, the main
effects and the interaction effects of the controllable factors are
evaluated as Table \ref{tbl:anova-example}.
% R command:
% dat <- read.table("tab1.txt",header=T)
% attach(dat)
% FA <- factor(FA)
% FB <- factor(FB)
% FC <- factor(FC)
% summary(aov(SN ~ FA+FB+FC+FA:FB+FA:FC+FB:FC))
%             Df Sum Sq Mean Sq F value Pr(>F)
% FA           1 461.35  461.35 18.8199 0.1442
% FB           1  45.90   45.90  1.8723 0.4018
% FC           1  91.27   91.27  3.7233 0.3044
% FA:FB        1  38.10   38.10  1.5541 0.4304
% FA:FC        1 278.17  278.17 11.3475 0.1837
% FB:FC        1  12.78   12.78  0.5214 0.6019
% Residuals    1  24.51   24.51               
%
% summary(aov(SN ~ FA+FB+FC+FA:FC))
%            Df Sum Sq Mean Sq F value  Pr(>F)  
%FA           1 461.35  461.35 18.3578 0.02336 *
%FB           1  45.90   45.90  1.8263 0.26944  
%FC           1  91.27   91.27  3.6319 0.15276  
%FA:FC        1 278.17  278.17 11.0689 0.04482 *
%Residuals    3  75.39   25.13                  
%---
%Signif. codes:  0 '***' 0.001 '**' 0.01 '*' 0.05 '.' 0.1 ' ' 1 
%
\begin{table*}
\begin{center}
\caption{Results of ANOVA for Table \ref{tbl:example-miyakawa-7.3}}
\begin{tabular}{ccr@{.}lr@{.}lr@{.}lr@{.}l}\hline
Factors & df & \multicolumn{2}{c}{Sum sq} & \multicolumn{2}{c}{Mean sq} & 
\multicolumn{2}{c}{$F$ value} & 
\multicolumn{2}{c}{$p$ value}\\ \hline
$\BA$          & 1 & 461&35 & 461&35 & 18&3578 & 0&02336$\ast$\\
$\BB$          & 1 &  45&90 &  45&90 &  1&8263 & 0&26944\\
$\BC$          & 1 &  91&27 &  91&27 &  3&6319 & 0&15276\\
$\BA\times\BC$ & 1 & 278&17 & 278&17 & 11&0689 & 0&04482$\ast$\\
Residuals      & 3 &  75&39 &  25&13 & \multicolumn{4}{c}{}\\ \hline
\end{tabular}
\label{tbl:anova-example}
\end{center}
\end{table*}
Note that we include the interaction effects $\BA\times \BB$ and
$\BB\times \BC$
into the residuals.
This result suggests the existence of the interaction effect $\BA\times \BC$
in addition to the main effects $\BA$ and $\BC$. Therefore an optimal
condition is considered as $\BA_2\BC_1$, and the controllable factor $\BB$ does
not have a significant influence on performance of the inspection. 

The above procedure is simple and useful in applications. However, it
seems that the result can be unreliable when 
the estimated standard errors of the standardized
SN ratios are unbalanced. This is caused by the fact that the standardized
SN ratio is only a function of the sample odds ratio, and therefore
the sample size
for each run is not appropriately taken into account in the
procedure. 
%To clarify the problem, multiply the sample sizes for 
%the runs No. 1, 3, 6 and 8 by 10 in Table \ref{tbl:imaginary-data}.
To clarify the problem, divide the sample sizes for the runs No. 5 and 6 by
4 and 2, respectively,  which yields Table \ref{tbl:imaginary-data}.
\begin{table*}
\begin{center}
\caption{Imaginary data set}
\begin{tabular}{|c|ccc|rrrr|}\hline
No. & A & B & C & \multicolumn{1}{|c}{$n_{11}$} &
 \multicolumn{1}{c}{$n_{12}$} & \multicolumn{1}{c}{$n_{21}$} &
 \multicolumn{1}{c|}{$n_{22}$}\\ \hline
1 & 1 & 1 & 1 & 27 & 13 & 11 & 9 \\ % 1.69763
2 & 1 & 1 & 2 & 25 & 15 & 3 & 17 \\
3 & 1 & 2 & 1 & 17 & 23 & 8 & 12 \\
4 & 1 & 2 & 2 & 15 & 25 & 2 & 18 \\
5 & 2 & 1 & 1 & 10 & 0 & 3 & 2 \\
6 & 2 & 1 & 2 & 19 & 1 & 5 & 5 \\
7 & 2 & 2 & 1 & 40 & 0 & 11 & 9 \\
8 & 2 & 2 & 2 & 37 & 3 & 10 & 10 \\ \hline
\end{tabular}
\label{tbl:imaginary-data}
\end{center}
\end{table*}
Of course, the result of ANOVA is
almost the same to Table \ref{tbl:anova-example} for this
situation. Note that the difference is only caused by the
modification (\ref{eqn:modification}).
However, it is obvious that the
relatively good contributions of the interaction effect $\BA\times \BC$ 
in the run 5 should be underestimate. Consequently, 
the $p$ value for the interaction effect $\BA\times \BC$ can increase.

In this paper, we consider inference for the main and the interaction
effects
of several controllable factors in two-valued input-output systems.
For the designed experiments with counts data, the theory of the generalized
linear models (McCullagh and Nelder, 1989) are frequently used.
See Hamada and Nelder (1997), Chapter 13 of Wu and Hamada (2000) or
Aoki and Takemura (2006) for example.
We also rely on the general theory of the generalized linear models 
and construct a general testing procedure
for various hypotheses of the interaction effects in Section 2. 
Note that our settings are not limited to the full factorial designs.
Considering aliasing relations carefully, fractional factorial designs
are also treated by our procedure. In particular, we focus on the
relation to the models for the high-dimensional contingency tables
in Section 3.
For example, as we will show, the data of 
Table \ref{tbl:example-miyakawa-7.3} can be treated as $2^5$ contingency
table, and the model $\BA\times \BC$ is shown to be equivalent to one
of the hierarchical models for the five-dimensional contingency tables.
In Section 4, we give some numerical examples and show the effectiveness
of our procedure.

\section{Conditional tests for the interaction effects of the multiple
controllable factors}
%In this section, we give a main result of this paper. For the purpose of
%illustration, we apply our procedure to the example of Table
%\ref{tbl:example-miyakawa-7.3} after give a general definition in
%Section 2.1. Though
%the $2^3$ full factorial design is used in the 
%example of Table \ref{tbl:example-miyakawa-7.3},
%our method is also used for various
%fractional factorial designs, which we consider in Section 2.2.

%\subsection{Theories and construction of tests}
Suppose there are $K$ two-valued input-output systems, each of which
is constructed by setting the levels of several controllable factors.
As we have seen in Section 1, 
the observation for each run is summarized as $2\times 2$ contingency table.
We write the observation for the $k$th run as
\begin{equation}
n_{11k}, n_{12k}, n_{21k}, n_{22k}.
\label{eqn:2x2-table}
\end{equation}
%which are assumed to be a realization of random variables $N_{11k},
%N_{12k}, N_{21k}, N_{22k}$, respectively. 
It is natural to suppose 
an independent binomial model for the observations.
We write the occurrence parameters as $p_{ijk},\ i,j = 1,2;\ k = 1,\ldots,K$, 
where $p_{i1k} + p_{i2k} = 1$ for $i = 1,2;\ k = 1,\ldots,K$. 
The likelihood function is written as
\[
\displaystyle\prod_{k = 1}^K 
\prod_{i = 1}^2\prod_{j = 1}^2
{n_{1\cdot k}\choose{n_{11k}}}
{n_{2\cdot k}\choose{n_{21k}}}
p_{ijk}^{n_{ijk}},
\]
where $n_{i\cdot k} = n_{i1k} + n_{i2k}$. In this paper, 
following the convention of the analysis of the contingency tables, 
we use the similar dot notations
to express various marginal totals of the observations.
For example, we write
\[
n_{ij\cdot} = \sum_{k = 1}^{K}n_{ijk},\ 
n_{i\cdot\cdot} = \sum_{j = 1}^2\sum_{k = 1}^{K}n_{ijk}
\]
and so on.

To express various models for the parameter $p_{ijk}$, we use the
theory of the generalized linear models. Since we know that the odds ratio
\[
\theta_k = \frac{p_{11k}p_{22k}}{p_{12k}p_{21k}}
\]
is a good measure for evaluating performance of each system, we
assume the structure
\[
\log\theta_k = \beta_0 + \beta_1x_{k1} + \cdots 
+ \beta_{\nu-1}x_{k\nu-1},\ k = 1,\ldots,K
\]
where $x_{k1},\ldots,x_{k\nu-1}$ are the covariates. We write
the $\nu$-dimensional parameter $\beta$ and the covariate matrix $X$
as
\begin{equation}
\beta = (\beta_0,\beta_1,\ldots,\beta_{\nu-1})'
\label{eqn:parameter-beta}
\end{equation}
and
\[
X = \left(
\begin{array}{cccc}
1 & x_{11} & \cdots & x_{1\nu-1}\\
\vdots & \vdots & \cdots & \vdots\\
1 & x_{K1} & \cdots & x_{K\nu-1}
\end{array}
\right) = 
\left(
\begin{array}{cccc}
\Bone_K & \Bx_1 & \cdots & \Bx_{\nu-1},
\end{array}
\right)
\]
where $\Bone_K = (1,\ldots,1)'$ is the $K$-dimensional column
vector. We also define the $K$-dimensional frequency vector
\[
\Bn = (n_{111},\ldots,n_{11K})'.
\]
Then the likelihood function is written as
\[\begin{array}{cl}
& \displaystyle\prod_{k = 1}^K 
\prod_{i = 1}^2\prod_{j = 1}^2
{n_{1\cdot k}\choose{n_{11k}}}
{n_{2\cdot k}\choose{n_{21k}}}
p_{ijk}^{n_{ijk}}\\
= & 
\displaystyle\prod_{k = 1}^K 
{n_{1\cdot k}\choose{n_{11k}}}
{n_{2\cdot k}\choose{n_{21k}}}
p_{12k}^{n_{1\cdot k}}
p_{22k}^{n_{2\cdot k}}
\left(
\frac{p_{21k}}{p_{22k}}
\right)^{n_{\cdot 1k}}
\exp(n_{11k}\log\theta_k)\\
= & \left[
\displaystyle\prod_{k = 1}^K 
{n_{1\cdot k}\choose{n_{11k}}}
{n_{2\cdot k}\choose{n_{21k}}}
p_{12k}^{n_{1\cdot k}}
p_{22k}^{n_{2\cdot k}}
\left(
\frac{p_{21k}}{p_{22k}}
\right)^{n_{\cdot 1k}}
\right]
\exp\left(\displaystyle\sum_{k = 1}^K 
n_{11k}\log\theta_k\right)\\
= & \left[
\displaystyle\prod_{k = 1}^K 
{n_{1\cdot k}\choose{n_{11k}}}
{n_{2\cdot k}\choose{n_{21k}}}
p_{12k}^{n_{1\cdot k}}
p_{22k}^{n_{2\cdot k}}
\left(
\frac{p_{21k}}{p_{22k}}
\right)^{n_{\cdot 1k}}
\right]
\exp\left(
\beta_0\Bone_K'\Bn + 
\displaystyle\sum_{j = 1}^{\nu - 1} 
\beta_j \Bx_j' \Bn
\right)\\
= & \left[
\displaystyle\prod_{k = 1}^K 
{n_{1\cdot k}\choose{n_{11k}}}
{n_{2\cdot k}\choose{n_{21k}}}
p_{12k}^{n_{1\cdot k}}
p_{22k}^{n_{2\cdot k}}
\left(
\frac{p_{21k}}{p_{22k}}
\right)^{n_{\cdot 1k}}
\right]
\exp\left(
\beta'X'\Bn
\right),
\end{array}
\]
which implies that 
the sufficient statistic for the parameter is 
$n_{i\cdot k}, n_{\cdot jk}, i,j = 1,2, k = 1,\ldots,K$, which are
the marginal totals of the $2\times 2$ table of (\ref{eqn:2x2-table}), 
and $X'\Bn = (\Bone_K'\Bn,\Bx_1'\Bn,\ldots,\Bx_{\nu-1}'\Bn)$.

Now we consider the covariate matrix $X$. In this paper
we consider the typical situation that the run sequence of
the experimental units is allocated to each row of an orthogonal
design matrix. For example, the run sequence of 
Table \ref{tbl:example-miyakawa-7.3} in Section 1 is allocated
to the $2^3$ full factorial design. 
We write the orthogonal design matrix where the run sequence of the
experimental units is allocated as $K\times p$ matrix $D$, where each
element is $+1$ or $-1$. 
For example of Table \ref{tbl:example-miyakawa-7.3},
we have
\begin{equation}
D = \left(\begin{array}{ccc}
1 & 1 & 1\\
1 & 1 & -1\\
1 & -1 & 1\\
1 & -1 & -1\\
-1 & 1 & 1\\
-1 & 1 & -1\\
-1 & -1 & 1\\
-1 & -1 & -1\\
\end{array}\right).
\label{eqn:2-3-full-factorial-D}
\end{equation}
In this paper, we only consider the situation
that each controllable factor has two levels.
For later use, we write $D = (d_{ij}) =
(\Bd_1,\ldots,\Bd_p)$ where $\Bd_j = (d_{1j},\ldots,d_{Kj}) \in
\{-1,+1\}^K$ is the $j$-th column vector of $D$. We also define a simple
convention
\[
 \Bd_{st} = (d_{1s}d_{1t},\ldots,d_{Ks}d_{Kt})'
\]
and
\[
 \Bd_{stu} = (d_{1s}d_{1t}d_{1u},\ldots,d_{Ks}d_{Kt}d_{Ku})'
\]
for $1 \leq s < t < u \leq p$. 

The matrix $X$ is constructed from the design matrix $D$ to reflect the
main effects of the controllable factors and their interactions which we intend to
measure. For example, a simple model which only includes the main
effects of each controllable factor is given as $X = (\Bone_K\ D)$. Of
course, we can consider more complicated models containing various
interaction effects. In particular, the saturated model includes $K$
parameters. When $K$ is a power of $2$, the covariate matrix $X$ for the
saturated model is the Hadamard matrix of order $K$. In the case of
$2^3$ full factorial design (\ref{eqn:2-3-full-factorial-D}), for
example, the covariate matrix for the saturated model is given as
\begin{equation}
 X = (\Bone_K\ \Bd_1\ \Bd_2\ \Bd_{12}\ \Bd_3\ \Bd_{13}\ \Bd_{23}\ \Bd_{123}).
\label{eqn:Hadamard-8}
\end{equation}
Since the saturated model cannot be tested, we consider an appropriate
submodel of the saturated model. For the purpose of illustration, we
again focus on the example of Table \ref{tbl:example-miyakawa-7.3}. 
Since the analysis in Section 1 implies the model of the two main
effects $\BA$, $\BC$ and the interaction effect $\BA\times \BC$, we treat
this model as the {\it null model} and consider significance tests. 
Hereafter, we write this model as $\BA\BC$ by the manner of the
hierarchical models. 

This null hypothesis can be described by the parameter $\beta$ as
follows. Permuting the columns of (\ref{eqn:Hadamard-8}), we partition
the covariate matrix $X$ of the saturated model as
\[
 \begin{array}{l}
  X = (X_0\ X_1),\\
  X_0 = (\Bone_K\ \Bd_1\ \Bd_3\ \Bd_{13}) = (\Bone_K\ \Bx_1\ \Bx_2\ \Bx_3),\\
  X_1 = (\Bd_2\ \Bd_{12}\ \Bd_{23}\  \Bd_{123}) = (\Bx_4\ \Bx_5\ \Bx_6\
   \Bx_7),\\
 \end{array}
\]
and consider the corresponding parameter $\beta =
(\beta_0,\beta_1,\ldots,\beta_7)$. Note that $\nu - 1 = 3$ in
(\ref{eqn:parameter-beta}) in this case. Then the null hypothesis is described as
\[
 \mbox{H}_0:\ \beta_{\nu} = \cdots = \beta_{K - 1} = 0.
\] 
Under the null hypothesis H$_0$, the nuisance parameters are $\beta_0,
\ldots, \beta_{\nu - 1}$ and the sufficient statistic for the nuisance
parameters is written as
\begin{equation}
 \{n_{i\cdot k}\},\ \{n_{\cdot jk}\},\ i,j = 1,2;\ k = 1,\ldots,K,\
 X_0'\Bn.
\label{eqn:sufficient-statistic}
\end{equation}
Then by the theory of the similar tests, we can consider the conditional
tests based on the conditional distribution of $\Bn$ given
(\ref{eqn:sufficient-statistic}).

Now we consider significance tests of null hypothesis H$_0$, against
various alternative hypothesis H$_1$. Again for the purpose of
illustration, we consider the example of Table
\ref{tbl:example-miyakawa-7.3}. In this case, an important alternative
is to test the effect of a single additional effect, the main effect of
$\BB$. This alternative hypothesis is written as
\begin{equation}
 \mbox{H}_1:\ \beta_{\nu} \neq 0,\ \beta_{\nu + 1} = \cdots = \beta_{K -
 1} = 0.
\label{eqn:H1}
\end{equation}
Or we can also consider the goodness-of-fit of the null hypothesis. In this
case, the alternative hypothesis is written as
\[
 \mbox{H}_1:\ (\beta_{\nu}  = \cdots = \beta_{K - 1}) \neq (0,\ldots,0).
\]
Depending on the alternative hypothesis, we can use appropriate test
statistic $T(\Bn)$. For the alternative hypothesis written as (\ref{eqn:H1}),
for example, a typical test statistic is a deviance function
\[
2\sum_{i,j,k}n_{ijk}\log\displaystyle\frac{\widetilde{n_{ijk}}}{\widehat{n_{ijk}}},
\]
where $\widehat{n_{ijk}}$ and $\widetilde{n_{ijk}}$ are
the fitted values under H$_0$ and H$_1$, respectively.
Note that $\{\widehat{n_{ijk}}\}$ and $\widetilde{n_{ijk}}$ are calculated
from the sufficient statistics under the hypothesis, i.e., 
\[
 \{n_{i\cdot k}\},\ \{n_{\cdot jk}\},\ i,j = 1,2;\ k = 1,\ldots,K,\
 X_0'\Bn
\]
and
\[
 \{n_{i\cdot k}\},\ \{n_{\cdot jk}\},\ i,j = 1,2;\ k = 1,\ldots,K,\
 X_0'\Bn,\ \Bx_{\nu}'\Bn,
\]
respectively.
In Section 4, we perform various tests for Table
\ref{tbl:example-miyakawa-7.3}.

Finally of this section, we give a brief remark on the case that
the design is fractional factorial designs. 
From the arguments above, it is obvious that our procedure is also
applicable for the case of fractional factorial designs.
All that we have to pay attention is the consideration on the 
aliasing relation when we construct a model including interaction
effects. To illustrate, we again consider Table \ref{tbl:example-miyakawa-7.3}.
Suppose there is another controllable factor $\BD$. If we only consider eight-run
design ($K = 8$), the main effect of $\BD$ has to be confounded to some
interaction
effect of $\BA,\BB,\BC$, i.e., $2^{4-1}$ fractional factorial design is
considered.
When we define $\BD = \BA\BC$, for example, at most only one of 
the main effect $\BD$ and the interaction effect $\BA\times\BC$ is estimable. 
Similarly, the interaction effects $\BA\times\BD$ and $\BC\times\BD$ are
also not estimable
when the main effects $\BC$ and $\BA$ exist, respectively.
Then the resolution of such design is III. 
On the other hand, if we define $\BD = \BA\BB\BC$, we can estimate all the
two-factor interaction effects, under the constraints that at most only one of
$(\BA\times\BB, \BC\times\BD)$, $(\BA\times\BC, \BB\times\BD)$ and
$(\BA\times\BD, \BB\times\BC)$ is included
in the model. The resolution of the design is IV. 
In Section 4, we consider such situations for Table
\ref{tbl:example-miyakawa-7.3}.

%\subsection{Note on the case of fractional factorial designs}

\section{Relation to the high-dimensional contingency tables}
In Section 2, we give a general procedure to describe models by the covariate
matrix $X$ and the parameter $\beta$, and to describe the sufficient statistic
under the models. As we have seen, statistical tests are based on some
discrepancy measure between the 
fitted values under the null and the alternative hypotheses, 
which are calculated from the sufficient statistics of the form
(\ref{eqn:sufficient-statistic}). In this section, we give an interpretation of
the sufficient statistics by considering the high-dimensional contingency
tables. The main concepts of the arguments in this section are 
first shown in the previous
work by one of the authors, Aoki and Takemura (2006). 

\paragraph*{Full factorial designs}
When the experiment is allocated to the full factorial design, there is
a direct correspondence to the high-dimensional contingency tables as
follows. Suppose the design is $2^p$ full factorial design, where $2^p =
K$. In this case, we can treat the observations $n_{ijk},\ i,j = 1,2;\ k
= 1,\ldots,K$ as if they are the frequencies of $2^{p+2}$ contingency
tables. To illustrate this point, we rewrite the cell indices of the
frequency as $n_{ija_1a_2\cdots a_p}$, where $i,j,a_1,\ldots,a_p = 1,2$.
Consequently, the observation of Table \ref{tbl:example-miyakawa-7.3},
i.e., the case of $p = 3$, is written as follows.
\begin{center}
\begin{tabular}{|c|ccc|rrrr|}\hline
No. & A & B & C & \multicolumn{1}{|c}{$n_{11}$} &
 \multicolumn{1}{c}{$n_{12}$} & \multicolumn{1}{c}{$n_{21}$} &
 \multicolumn{1}{c|}{$n_{22}$}\\ \hline
1 & 1 & 1 & 1 & $n_{11111}$ & $n_{12111}$ & $n_{21111}$ & $n_{22111}$ \\ 
2 & 1 & 1 & 2 & $n_{11112}$ & $n_{12112}$ & $n_{21112}$ & $n_{22112}$ \\ 
3 & 1 & 2 & 1 & $n_{11121}$ & $n_{12121}$ & $n_{21121}$ & $n_{22121}$ \\ 
4 & 1 & 2 & 2 & $n_{11122}$ & $n_{12122}$ & $n_{21122}$ & $n_{22122}$ \\ 
5 & 2 & 1 & 1 & $n_{11211}$ & $n_{12211}$ & $n_{21211}$ & $n_{22211}$ \\ 
6 & 2 & 1 & 2 & $n_{11212}$ & $n_{12212}$ & $n_{21212}$ & $n_{22212}$ \\ 
7 & 2 & 2 & 1 & $n_{11221}$ & $n_{12221}$ & $n_{21221}$ & $n_{22221}$ \\ 
8 & 2 & 2 & 2 & $n_{11222}$ & $n_{12222}$ & $n_{21222}$ & $n_{22222}$ \\ \hline
\end{tabular}
\end{center}
Now we focus on the sufficient statistics under various models in this
notation. For notation convenience, we use $a,b,c$ instead of
$a_1,a_2,a_3$ here. We have already seen that the marginal totals
\[
 \{n_{i\cdot abc}\},\ \{n_{\cdot jabc}\}
\]
are included in the sufficient statistic for every model. Under the main
effect model $\BA/\BB/\BC$, the sufficient statistic is given as
\[
 \{n_{i\cdot abc}\},\ \{n_{\cdot jabc}\},\ \{n_{ija\cdot\cdot}\},\
 \{n_{ij\cdot b\cdot}\},\ \{n_{ij\cdot\cdot c}\}.
\]
We know that this is the sufficient statistics of the hierarchical model 
\[
 M\BA\BB\BC/y\BA\BB\BC/My\BA/My\BB/My\BC
\]
in the five-way contingency tables. 
On the other hand, under the model of
$\BA\BC$, i.e., two main effects $\BA$, $\BC$ and the interaction effect
$\BA\times \BC$, the sufficient statistic is given as
\[
 \{n_{i\cdot abc}\},\ \{n_{\cdot jabc}\},\ \{n_{ija\cdot c}\},
\]
which is the sufficient statistic of the hierarchical model
\[
 M\BA\BB\BC/y\BA\BB\BC/My\BA\BC
\]
in the five-way contingency tables. In the same way, all the hierarchical
models of the effects of the controllable factors $\BA,\BB,\BC$ 
can be characterized as the 
corresponding hierarchical models in the $2^5$ contingency tables. We
give the relations in Table \ref{tbl:model-8-full-factorial}.
\begin{table*}[htbp]
\begin{center}
\caption{Hierarchical
models of the effects of the controllable factors $\BA,\BB,\BC$ and their 
corresponding hierarchical models in $2^5$ contingency tables ($2^3$
 full factorial design)}
\begin{tabular}{ll}\hline
Models for $\BA,\BB,\BC$ & Models for $M, y, \BA,\BB,\BC$ \\ \hline
$\BA$ & $M\BA\BB\BC/y\BA\BB\BC/My\BA$\\
$\BA/\BB$ & $M\BA\BB\BC/y\BA\BB\BC/My\BA/My\BB$\\
$\BA\BB$ & $M\BA\BB\BC/y\BA\BB\BC/My\BA\BB$\\
$\BA/\BB/\BC$ & $M\BA\BB\BC/y\BA\BB\BC/My\BA/My\BB/My\BC$\\
$\BA\BB/\BC$ & $M\BA\BB\BC/y\BA\BB\BC/My\BA\BB/My\BC$\\
$\BA\BB/\BA\BC$ & $M\BA\BB\BC/y\BA\BB\BC/My\BA\BB/My\BA\BC$\\
$\BA\BB/\BA\BC/\BB\BC$ &
 $M\BA\BB\BC/y\BA\BB\BC/My\BA\BB/My\BA\BC/My\BB\BC$\\ \hline
\end{tabular}
\label{tbl:model-8-full-factorial}
\end{center}
\end{table*}
Similarly, we can consider full factorial designs with 16 runs, 32 runs
and so on. All the hierarchical models of full factorial designs with $2^p$ runs
can be treated as the corresponding hierarchical models in $2^{p+2}$ contingency
tables.

\paragraph*{Fractional factorial designs}
Next we consider the case of fractional factorial design. 
Again we consider the design with 8 runs for illustration. 
In the case of $p = 4$, it is natural to define the aliasing relation as
$\BD = \BA\BB\BC$ since this gives the design of resolution IV. The
observation is written as follows.
\begin{center}
\begin{tabular}{|c|cccc|rrrr|}\hline
No. & A & B & C & D & \multicolumn{1}{|c}{$n_{11}$} &
 \multicolumn{1}{c}{$n_{12}$} & \multicolumn{1}{c}{$n_{21}$} &
 \multicolumn{1}{c|}{$n_{22}$}\\ \hline
1 & 1 & 1 & 1 & 1 & $n_{11111}$ & $n_{12111}$ & $n_{21111}$ & $n_{22111}$ \\ 
2 & 1 & 1 & 2 & 2 & $n_{11112}$ & $n_{12112}$ & $n_{21112}$ & $n_{22112}$ \\ 
3 & 1 & 2 & 1 & 2 & $n_{11121}$ & $n_{12121}$ & $n_{21121}$ & $n_{22121}$ \\ 
4 & 1 & 2 & 2 & 1 & $n_{11122}$ & $n_{12122}$ & $n_{21122}$ & $n_{22122}$ \\ 
5 & 2 & 1 & 1 & 2 & $n_{11211}$ & $n_{12211}$ & $n_{21211}$ & $n_{22211}$ \\ 
6 & 2 & 1 & 2 & 1 & $n_{11212}$ & $n_{12212}$ & $n_{21212}$ & $n_{22212}$ \\ 
7 & 2 & 2 & 1 & 1 & $n_{11221}$ & $n_{12221}$ & $n_{21221}$ & $n_{22221}$ \\ 
8 & 2 & 2 & 2 & 2 & $n_{11222}$ & $n_{12222}$ & $n_{21222}$ &
 $n_{22222}$ \\ \hline
\end{tabular}
\end{center}
In this case, because of the relation $\BD = \BA\BB\BC$, some
hierarchical models of the controllable factors $\BA,\BB,\BC,\BD$ does not
have a corresponding hierarchical model in $2^5$ contingency tables. For
example, the sufficient statistic for the main effect model
$\BA/\BB/\BC/\BD$ is written as follows.
\[\begin{array}{l}
 \{n_{i\cdot abc}\},\ \{n_{\cdot jabc}\},\ \{n_{ija\cdot\cdot}\},\
 \{n_{ij\cdot b\cdot}\},\ \{n_{ij\cdot\cdot c}\},\\
\{n_{ij111} + n_{ij122} + n_{ij212} + n_{ij221}\},\ 
\{n_{ij112} + n_{ij121} + n_{ij211} + n_{ij222}\},
\end{array}\]
which does not correspond to the 
sufficient statistic for any hierarchical model 
in $2^5$ contingency tables. 
Similarly, the sufficient statistic for 
the model $\BA\BB/\BC/\BD$, i.e., the model of the four main effects 
and the interaction effect $\BA\times \BB$, is written as
\[
\begin{array}{l}
 \{n_{i\cdot abc}\},\ \{n_{\cdot jabc}\},\ \{n_{ijab\cdot}\},\
\{n_{ij\cdot\cdot c}\},\\
\{n_{ij111} + n_{ij122} + n_{ij212} + n_{ij221}\},\ 
\{n_{ij112} + n_{ij121} + n_{ij211} + n_{ij222}\},
\end{array}\]
which again does not correspond to the 
sufficient statistic for any hierarchical model 
in $2^5$ contingency tables. 
Note here that the set $\{\Bd_{12}'\Bn, \Bd_1'\Bn, \Bd_2'\Bn\}$, i.e.,
\[
\{n_{ija\cdot\cdot}\},\ \{n_{ij\cdot b\cdot}\},\ 
\{n_{ij11\cdot}+n_{ij22\cdot}\},\ \{n_{ij12\cdot}+n_{ij21\cdot}\},
\]
is simply written as $\{n_{ijab\cdot}\}$ by the relation
\begin{equation}
n_{ijab\cdot} = \frac{n_{ija\cdot\cdot}+n_{ij\cdot b\cdot} - (n_{ijab^*\cdot} + n_{ija^*b\cdot})}{2}
\label{eqn:relation}
\end{equation}
where $\{a,a^*\}$ and $\{b,b^*\}$ are the distinct indices, respectively.
We see that only the saturated model $\BA\BB/\BA\BC/\BB\BC/\BD$
among the models including the main effect of $\BD$ has a 
corresponding hierarchical models of $2^5$ contingency tables. 

We also consider the case of $p=5,6$. If there are $5$ controllable factors, it is natural to
define the aliasing relation as $\BD = \BA\BB, \BE = \BA\BC$, which yields the design
of resolution III. In this case, 
the sufficient statistic for the main effect 
model $\BA/\BB/\BC/\BD/\BE$ is written as
\[
\begin{array}{l}
 \{n_{i\cdot abc}\},\ \{n_{\cdot jabc}\},\ \{n_{ijab\cdot}\},\ \{n_{ija\cdot c}\},
\end{array}
\]
which is the sufficient statistic for the hierarchical model $M\BA\BB\BC/y\BA\BB\BC/My\BA\BB/My\BA\BC$
of $2^5$ contingency tables. Similarly, consider the models containing interaction effect.
From the aliasing relation, the estimable interaction is $\BB\times \BC$, $\BB\times \BE$ or $\BC\times \BD$,
where the later two are also confounded.
We see that 
the sufficient statistic for the model 
$\BA/\BB\BC/\BD/\BE$ is written as
\begin{equation}
\begin{array}{l}
 \{n_{i\cdot abc}\},\ \{n_{\cdot jabc}\},\ \{n_{ijab\cdot}\},\ \{n_{ija\cdot c}\}\ \{n_{ij\cdot bc}\},
\end{array}
\label{eqn:A-BC-D-E-model}
\end{equation}
which is the sufficient statistic for the hierarchical model $M\BA\BB\BC/y\BA\BB\BC/My\BA\BB/My\BA\BC/My\BB\BC$
of $2^5$ contingency tables.
On the other hand, 
the sufficient statistic for the model 
$\BA/\BB\BE/\BC/\BD$ is written as
\[
\begin{array}{l}
 \{n_{i\cdot abc}\},\ \{n_{\cdot jabc}\},\ \{n_{ijab\cdot}\},\\
\{n_{ij111} + n_{ij122} + n_{ij212} + n_{ij221}\},\ 
\{n_{ij112} + n_{ij121} + n_{ij211} + n_{ij222}\},
\end{array}
\]
which does not correspond to the 
sufficient statistic for any hierarchical model in $2^5$ contingency tables. 
Finally, in the case of $p = 6$, consider the design defined as $\BD = \BA\BB, \BE = \BA\BC,
\BF = \BB\BC$. In this case, the sufficient statistic for 
the main effect model $\BA/\BB/\BC/\BD/\BE/\BF$ is written as (\ref{eqn:A-BC-D-E-model}), 
which is also the sufficient statistic for the hierarchical model $M\BA\BB\BC/y\BA\BB\BC/My\BA\BB/My\BA\BC/My\BB\BC$
of $2^5$ contingency tables.

In the same way, we can consider the fractional factorial designs with 16 runs, 32 runs and so on.
The arguments and results are almost similar. For example, if some controllable factor $\BF$ is
defined as $\BF = \BA\BB\BC$, the sufficient statistic for the model containing the main effect $\BF$
includes the marginal total of the type that some three-way marginal tables $\{n_{ijk}\}$ is written as
\[
n_{111}+n_{112}+n_{212}+n_{221},\ n_{112}+n_{121}+n_{211}+n_{222}.
\]
We see that, if all the two-way marginal tables $\{n_{ij\cdot}\}, \{n_{i\cdot k}\}, \{n_{\cdot jk}\}$ are 
given, then all the three-way
marginal totals $\{n_{ijk}\}$ can be derived from the similar relation to (\ref{eqn:relation})
such as
\[
\begin{array}{cl}
n_{111} = & \displaystyle\frac{1}{8}\left[3(n_{11\cdot}+n_{1\cdot 1}+n_{\cdot 11}) + (n_{22\cdot}+n_{2\cdot 2}+n_{\cdot 22})\right.\\
& \hspace*{10mm}-(n_{111}+n_{122}+n_{212}+n_{221})-3(n_{112}+n_{121}+n_{211}+n_{222})\left.\right].
\end{array}
\]

\section{Numerical examples}
In this section we investigate the data of Table
\ref{tbl:example-miyakawa-7.3}.
As we have seen in Section 1, a simple ANOVA analysis suggests the
interaction effect $\BA\times\BC$.
Therefore we perform the statistical test for the null model
$\BA/\BB/\BC$ against the alternative model
$\BA\BC/\BB$. As we have seen in Section 3, these hypotheses are
equivalent to the models of $2^5$
contingency tables, $M\BA\BB\BC/y\BA\BB\BC/My\BA/My\BB/My\BC$ (null model) and 
$M\BA\BB\BC/y\BA\BB\BC/My\BA\BC/My\BB$ (alternative model). The fitted
values for these models are given
in Table \ref{tbl:fitted-values}.

\begin{table*}
\begin{center}
\caption{Fitted values of Table \ref{tbl:example-miyakawa-7.3} under the
 null model $\BA/\BB/\BC$ (left)
and the alternative model $\BA\BC/\BB$ (right)}
\begin{tabular}{|c|ccc|rrrr|rrrr|}\hline
 & & & & \multicolumn{4}{|c|}{null model} &
 \multicolumn{4}{|c|}{alternative model}\\ 
No. & A & B & C & \multicolumn{1}{|c}{$n_{11}$} &
 \multicolumn{1}{c}{$n_{12}$} & \multicolumn{1}{c}{$n_{21}$} &
 \multicolumn{1}{c|}{$n_{22}$} & \multicolumn{1}{|c}{$n_{11}$} &
 \multicolumn{1}{c}{$n_{12}$} & \multicolumn{1}{c}{$n_{21}$} &
 \multicolumn{1}{c|}{$n_{22}$}\\ \hline
1 & 1 & 1 & 1 & 27.7 &12.3 &10.3 & 9.7  &   26.7  &  13.3  &  11.3  &
 8.7\\	  
2 & 1 & 1 & 2 & 24.1 &15.9 & 3.9 &16.1  &   25.2  &  14.8  &   2.8  &
 17.2\\	  
3 & 1 & 2 & 1 & 19.0 &21.0 & 6.0 &14.0  &   17.3  &  22.7  &   7.7  &
 12.3\\	  
4 & 1 & 2 & 2 & 13.2 &26.8 & 3.8 &16.2  &   14.8  &  25.2  &   2.2  &
 17.8\\	  
5 & 2 & 1 & 1 & 39.0 & 1.0 &13.0 & 7.0  &   40.0  &   0.0  &  12.0  &
 8.0\\	  
6 & 2 & 1 & 2 & 39.2 & 0.8 & 8.8 &11.2  &   38.1  &   1.9  &   9.9  &
 10.1\\	  
7 & 2 & 2 & 1 & 38.3 & 1.7 &12.7 & 7.3  &   40.0  &   0.0  &  11.0  &
 9.0\\	  
8 & 2 & 2 & 2 & 38.5 & 1.5 & 8.5 &11.5  &   36.9  &   3.1  &  10.1  &
 9.9\\ \hline
\end{tabular}
\end{center}
\label{tbl:fitted-values}
\end{table*}
The likelihood statistic is calculated as $13.06$ with degree of freedom
$1$. Therefore
we have a conclusion that the null model is rejected ($p = 0.000301$) and
the interaction effect
$\BA\times \BC$ is statistically significant. To show the efficacy of
our procedure, we also give 
an imaginary data set of Table \ref{tbl:imaginary-data}. The fitted
values are given 
in Table \ref{tbl:fitted-values-imaginary} in this case, and the
likelihood statistic is $11.56$. Though this result also suggests the
significance of the interaction effect $\BA\times \BC$, the $p$ value 
increases to $p = 0.000674$. 
\begin{table*}
\begin{center}
\caption{Fitted values of Table \ref{tbl:imaginary-data} under the null model $\BA/\BB/\BC$ (left)
and the alternative model $\BA\BC/\BB$ (right)}
\begin{tabular}{|c|ccc|rrrr|rrrr|}\hline
 & & & & \multicolumn{4}{|c|}{null model} & \multicolumn{4}{|c|}{alternative model}\\ 
No. & A & B & C & \multicolumn{1}{|c}{$n_{11}$} &
 \multicolumn{1}{c}{$n_{12}$} & \multicolumn{1}{c}{$n_{21}$} &
 \multicolumn{1}{c|}{$n_{22}$} & \multicolumn{1}{|c}{$n_{11}$} &
 \multicolumn{1}{c}{$n_{12}$} & \multicolumn{1}{c}{$n_{21}$} &
 \multicolumn{1}{c|}{$n_{22}$}\\ \hline
1 & 1 & 1 & 1 &  27.2  &  12.8  &  10.8  &   9.2  &        26.7 &   13.3 &   11.3 &    8.7\\	    
2 & 1 & 1 & 2 &  24.4  &  15.6  &   3.6  &  16.4  &	    25.3 &   14.7 &    2.7 &   17.3\\	    
3 & 1 & 2 & 1 &  19.1  &  20.9  &   5.9  &  14.1  &	    17.3 &   22.7 &    7.7 &   12.3\\	    
4 & 1 & 2 & 2 &  13.3  &  26.7  &   3.7  &  16.3  &	    14.7 &   25.3 &    2.3 &   17.7\\	    
5 & 2 & 1 & 1 &   9.7  &   0.3  &   3.3  &   1.7  &	    10.0 &    0.0 &    3.0 &    2.0\\	    
6 & 2 & 1 & 2 &  19.6  &   0.4  &   4.4  &   5.6  &	    19.1 &    0.9 &    4.9 &    5.1\\	    
7 & 2 & 2 & 1 &  38.0  &   2.0  &  13.0  &   7.0  &	    40.0 &    0.0 &   11.0 &    9.0\\	    
8 & 2 & 2 & 2 &  38.7  &   1.3  &   8.3  &  11.7  &	    36.9 &    3.1 &   10.1 &    9.9\\ \hline
\end{tabular}
\end{center}
\label{tbl:fitted-values-imaginary}
\end{table*}

\section{Discussion}
In this paper, we give a general testing procedure to investigate the
main and the interaction effects of the controllable factors in the
two-valued input-output systems. For this type of data set, simple ANOVA
model for the standardized SN ratio is widely used, which is a proposal
of Taguchi method. However, since the standard SN ratio is only a
function of the sample odds ratio, we cannot evaluate the influence of
the sample size of the data appropriately in the simple ANOVA model.
In contrast, since our method is based on the conditional distribution
for the various null models, sample size is considered in $p$ values.

Note that we do not deny the ANOVA model completely. It is
unquestionable that the simple ANOVA model is useful in
applications. However, it seems that the ANOVA model for the standard SN
ratio is introduced heuristically. We believe that we have to
investigate the theoretical validity for the Taguchi method carefully, and
modify it as the need arises. We think that our procedure in this paper
makes some contribution in this fields.

Though we only give an illustration of the likelihood ratio
test based on the asymptotic $\chi^2$ distribution in Section 4, various
exact procedures or Monte Carlo procedures are also possible. See
Agresti (1992) for the exact tests and  Aoki and Takemura (2006) for the
Markov chain Monte Carlo tests, for example.

\end{document}